\documentclass[aps,prb,twocolumn,amsmath,amssymb]{revtex4}
\usepackage{graphicx}
\usepackage{dcolumn}
\usepackage{bm}
\newcommand{\ds}{\displaystyle}

\begin{document}
\title{Size-dependent melting: Numerical calculations of the phonon
spectrum}

\author{Kai Kang}
\affiliation{School of Physics, Peking University, Beijing 100871,
China}
\author{Shaojing Qin}
\affiliation{Institute of Theoretical Physics, Chinese Academy of
Sciences, P.O. Box 2735, Beijing 100080, China}
\author{Chuilin Wang}
\affiliation{China Center of Advanced Science and Technology, P. 0.
Box 8730, Beijing 100080, China}

\date{\today}
\begin{abstract}

In order to clarify the relationship between the phonon spectra of
nanoparticles and their melting temperature, we studied in detail
the size-dependent low energy vibration modes. A minimum model with
atoms on a lattice and harmonic potentials for neighboring atoms is
used to reveal a general behavior. By calculating the phonon spectra
for a series of nanoparticles of two lattice types in different
sizes, we found that density of low energy modes increases as the
size of nanoparticles decreases, and this density increasing causes
decreasing of melting temperature. Size-dependent behavior of the
phonon spectra accounts for typical properties of surface-premelting
and irregular melting temperature on fine scales. These results show
that our minimum model captures main physics of nanoparticles.
Therefore, more physical characteristics for nanoparticles of
certain types can be given by phonons and microscopic potential
models.

\end{abstract}

\maketitle

\section{INTRODUCTION}
\label{sec:intro}

The physical properties of nanosized material have attracted
considerable attention because of its technological importance as
well as its fundamental interest for theoretical study. One
important thermodynamic property is its size-dependent melting point
depression, which, in fact, has been studied for decades. This
unusual behavior of nanoparticles can lead to develop new features
of materials with wide applications
\cite{STjong:2004,MZhang:2000,MZhao:2004}.

Experimental studies show that (with a few exceptions), for free
surface nanoparticles, the melting temperature decreases with
decreasing size, and has irregular variations on a fine scale
\cite{FBaletto:2005,CWang:2005,MTakagi:1954}. Many theoretical
models have been employed to investigate this unusual property. Of
these various models we can classify them into many categories,
thermodynamical models
\cite{RCouchman:1979,PPawlow:1909,CWronski:1967,HSakai:1996,
RVanfleet:1995}, latent heat model \cite{QJiang:1999,QJiang:2002},
the surface-phonon instability model
\cite{MWautelet:1995,MWautelet:1990,MWautelet:1991}, the liquid drop
model \cite{KNanda:2002,KNanda:2006}, and bond energy model
\cite{WQi:2003,WQi:2004}. Most of these models give the result that,
the nanoparticle melting temperature $T_{mn}$ relative to the bulk
melting temperature $T_{mb}$ decreases linearly with size in a
general form such as $T_{mn}/T_{mb}=1-A/D$, where $D$ is the
nanoparticle diameter and $A$ is a constant depending upon the
modeling parameters. Some other models have a nonlinear form to fit
better for smaller sizes ($D<10$nm) \cite{QJiang:1999,
ASafaei:2008}.

The melting properties of two-dimensional Penrose lattices have been
studied by Huang et.~al, who made use of the phonon spectrum and the
Lindemann melting criterion \cite{XHuang:1992}. It showed that the
boundary atoms always have larger atomic mean-squared thermal
vibration amplitudes, which means lower melting temperature. We will
see that a similar behavior also appears in three-dimensional
lattices in our calculation.

In this paper, we study the relation between nanoparticle melting
temperature and nanoparticle size by means of the phonon spectrum,
and we intend to reveal the melting properties of nanoparticles from
a microscopic model. With a statistical melting criterion similar to
the Lindemann criterion, we found that the linear depression of
nanoparticle melting temperature with decreasing size and the
irregular variations on a fine scale was given in our numerical
calculation by a very simple model.

\section{MODEL}

Two three-dimensional lattice types are considered in this paper.
For simplicity, we only calculate near-spherical nanoparticles in
this paper. To construct such nanoparticles, first, we create the
simple cubic (SC) or face-centered cubic (FCC) lattices with $n$
lattice points on the edge. Then cut an inscribed sphere in the
corresponding lattice type. The lattice points on and inside the
sphere are what we count in the calculation (actually there are no
lattice points on the sphere for even $n$ SC lattice). Assume that
there is one atom at each lattice point, which is the equilibrium
position of the atom, and the surface is free. So we do not have to
take surface reconstruction and relaxation into account in this
calculation.

We directly calculate the phonon modes in the harmonic approximation
for the two lattices of different sizes (different $n$). The
interatomic potentials that we use is the spring potential
\begin{eqnarray}
V&=&\frac{1}{2}K_1\sum _{\langle ij
\rangle}(|\bm{r}_0^i+\bm{u}^i-\bm{r}_0^j-\bm{u}^j|-|\bm{r}_0^i-\bm{r}_0^j|)^2\nonumber\\
& &+\frac{1}{2}K_2\sum_{\{ij\}}
(|\bm{r}_0^i+\bm{u}^i-\bm{r}_0^j-\bm{u}^j|-|\bm{r}_0^i-\bm{r}_0^j|)^2
\end{eqnarray}
where $\bm{r}_0^i$ is the position of the $i$th lattice point and
$\bm{u}^i$ is the displacement of the atom from its equilibrium
position $\bm{r}_0^i$; $\langle ij \rangle$ means that the sum is
taken over the nearest neighbors and $\{ij\}$ means the next-nearest
neighbors; $K_1$ and $K_2$ are spring constants for the nearest and
next-nearest interactions respectively. This potential ensures that
the structure of nanoparticle we consider is at least one of the
lowest energy structures.

Within the harmonic approximation, we can deduce that the atomic
mean-squared thermal vibration amplitude $\langle u_{i\alpha} ^2
\rangle$ is
\begin{equation}
\langle u_{i\alpha} ^2 \rangle = \sum _k \frac{e_{i\alpha,k} ^2}{m_i
\omega _k} (\frac{1}{e^{\omega_k/T}-1}+\frac{1}{2}) \label{eq:u2}
\end{equation}
where $i = 1, 2, \ldots ,N$ and $N$ is the number of atoms of the
nanoparticle; $\alpha = x, y, z$; $m_i$ is the mass of the $i$th
atom; $\omega _k$ is the $k$th eigenfrequency, and $e_{i\alpha,k}$
is the component of the eigenvector corresponding to the $k$th
eigenfrequency. We set $\hbar=k_B=1$ here. $\langle$ $\rangle$ means
grand canonical ensemble average.

In bulk materials, the Lindemann criterion \cite{FLindemann:1910} is
commonly used. It states that a material melts at the temperature at
which the amplitude of thermal vibration exceeds a certain critical
fraction of the interatomic distance. Experiments and simulations
show that its critical value is around 0.10--0.15 in units of the
interatomic distance. Lawson showed that the overall accuracy of the
Lindemann criterion for the materials he considered is not better
than about 30\%, and he gave an improved Lindemann criterion
\cite{ALawson:2001}. For irregular nanoparticles, the
distance-fluctuation criterion is introduced
\cite{REtters:1977,RBerry:1988}. It is based on the fluctuation of
the distance between pairs of atoms, while the Lindemann criterion
is based on the fluctuation of individual atoms relative to their
average position. And it has a problem that it depends on the
duration of the simulation run \cite{DFrantz:1995,DFrantz:2001}.

It can be seen from the expression of $\langle u_{i\alpha} ^2
\rangle$ that, when the atoms all have the same mass $m_i\equiv m$,
the total mean-squared atomic thermal vibration amplitude $\langle
u^2\rangle_{tot}=\ds\sum_{i,\alpha}\langle u_{i\alpha} ^2 \rangle$
is determined only by the eigenfrequencies except the mass $m$,
because of the orthonormality of the eigenvectors
$\ds\sum_{i,\alpha}e_{i\alpha,k}^2=1$, i.e.~
\begin{equation}\langle
u^2\rangle_{tot}=\frac{1}{m}\sum_k\frac{1}{\omega_k}(\frac{1}{e^{\omega_k/T}-1}+\frac{1}{2}).
\end{equation}
In this case, we can directly use the Lindemann criterion in our
model and do not need to calculate the eigenvectors at all.

But when the atoms do not have the same mass, we have to and
actually we can easily calculate each $\langle u_{i\alpha} ^2
\rangle$ numerically here. Therefor it is reasonable to use a
statistical criterion like the Lindemann criterion: \textit{if more
than fifty percent of the $\sqrt{\langle u_{i\alpha} ^2 \rangle}$
exceed half of the interatomic distance, the melting happens}. The
``fifty" and ``half" are not particularly selected and can be tuned.
Compared to the Lindemann criterion, the effect of using these two
figures together may be too large. But obviously they can be tuned
to be consistent with the Lindemann criterion. And because the model
we use here is so simple and we just focus on qualitative
description, there should be no problem to use these two figures in
the moment. We will show that, with this statistical criterion, the
linear depression of nanoparticle melting temperature with
decreasing size and the irregular variations on a fine scale are
reproduced in our model and a crosspoint in the phonon modes and
related results are discovered when we consider $\langle u_{i\alpha}
^2 \rangle$ statistically, although we only consider the case that
all the atoms have the same mass.

Actually, the eigenfrequencies depend only on the mass $m_i$ and the
spring constant $K_1,K_2$ in our model. So each $\langle u_{i\alpha}
^2 \rangle$ at a given temperature is completely determined when
$m_i$ and $K_1,K_2$ have been chosen. Different lattice spacing $a$
has different melting temperature $T_{mn}$ because of its appearance
as equilibrium interatomic distance in the criterion.

\section{RESULTS AND DISCUSSION}
In the present calculation, we only consider the case of monoatomic
nanoparticles. The SC type lattice has the nearest-neighbor
interaction as well as the next-nearest-neighbor interactions, while
FCC type only has the nearest-neighbor interaction. We set $m_i
\equiv m=1,K_1=3,K_2=2,a=1$ for SC and $m_i \equiv
m=1,K_1=9,K_2=0,a=\sqrt[3]{4}$ for FCC. We choose $a=1$ for SC and
$a=\sqrt[3]{4}$ for FCC so that they have the same number density.

\begin{figure}
\includegraphics[width=8cm]{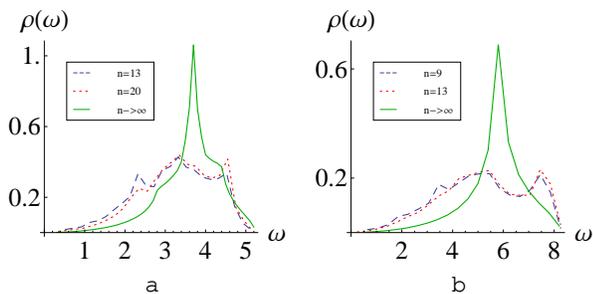}
\caption{(Color online)Density of phonon modes for different sizes.
(a) SC lattices. The dashed (blue) line is for $n=13$, the dotted
(red) line is for $n=20$, and the solid (green) line is for
$n\rightarrow\infty$. (b) FCC lattices. The dashed (blue) line is
for $n=9$, the dotted (red) line is for $n=13$, and the solid
(green) line is for $n\rightarrow\infty$. $n$ is the number of
lattice points on the edge of the cubic lattice, from which we cut
the inscribed sphere.} \label{fig:dos}
\end{figure}

The density of phonon modes for lattices in two different finite
sizes along with their thermodynamic limit are shown in
Fig.~\ref{fig:dos} for SC and FCC respectively. When the size of
nanoparticle becomes larger, the profile of the density of phonon
modes becomes smoother.

For the two finite sizes of SC lattices, there are both three peaks
at $\omega \approx$ 2.4, 3.4, and 4.6. The peak at $\omega \approx
2.4$ lowers down with increasing size, and it becomes a shoulder at
a higher frequency $\omega \approx 2.8$ ultimately in the
thermodynamic limit. So does the peak at $\omega \approx 4.6$, but
it increases with increasing size and the shoulder in the
thermodynamic limit is at a lower frequency $\omega \approx 4.4$.
The peak at $\omega \approx 3.4$ for finite sizes becomes the only
peak at a higher frequency $\omega \approx 3.7$ in the thermodynamic
limit. This peak in the thermodynamic limit arises from the symmetry
that leads to a large number of degenerate eigenfrequencies at this
value. And our calculation show that when the symmetry breaks, the
degeneracy is lifted and peaks beside it appear in finite sizes. The
situation for FCC lattices is similar. But there are no shoulders in
the thermodynamic limit and the peak at $\omega\approx3.5$ in the
smaller finite size disappear in the larger finite size while the
peak at $\omega\approx7.5$ does not change much for FCC.

The density of phonon modes for nanoparticles of those sizes, which
are smaller than the smaller finite size ($n=13$ for SC and $n=9$
for FCC) appears in Fig.~\ref{fig:dos}, begins to have zigzag-like
profiles and to fluctuate more and more strongly when the size gets
smaller and smaller.

To see which phonon modes contribute to the melting, we pick out
those $\langle u_{i\alpha} ^2 \rangle$ that meet our criterion and
plot the mean contribution from each
$[\omega,\omega+\mathrm{d}\omega]$ to $\langle u_{i\alpha} ^2
\rangle$ versus $\omega$ at the nanoparticle melting temperature
determined by our model. The mean contribution from each
$[\omega,\omega+\mathrm{d}\omega]$ to $\langle u_{i\alpha} ^2
\rangle$ is defined as
\begin{equation}
P(\omega)\equiv\frac{1}{N_m}{\sum_{i,\alpha}}'\frac{\langle
u_{i\alpha} ^2 \rangle_{\mathrm{d}\omega}}{\langle u_{i\alpha} ^2
\rangle}
\end{equation}
where $\sum'$ means the sum is taken over those $\langle u_{i\alpha}
^2 \rangle$ that meet our criterion, and
\begin{equation}
\langle u_{i\alpha} ^2 \rangle_{\mathrm{d}\omega}\equiv{\sum_{k}}
'\frac{e_{i\alpha,k} ^2}{m_i \omega _k}
(\frac{1}{e^{\omega_k/T}-1}+\frac{1}{2})
\end{equation}
means the sum of $k$ in the expression of $\langle u_{i\alpha} ^2
\rangle$ (Eq.~\ref{eq:u2}) is taken only over those $k$ that are in
the range of $\omega_k\in[\omega,\omega+\mathrm{d}\omega]$, and
$N_m$ means the number of those $\langle u_{i\alpha} ^2 \rangle$. We
choose $\mathrm{d}\omega=0.2$ for SC and $\mathrm{d}\omega=0.4$ for
FCC.

\begin{figure}
\includegraphics[width=8cm]{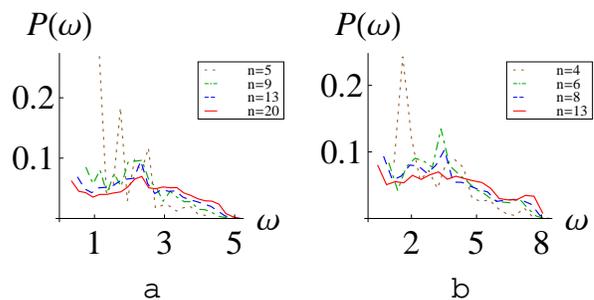}
\caption{(Color online)Mean contributions from different
eigenfrequencies to the melting. (a) SC lattices. The dotted (brown)
line is for $n=5$, the dotdashed (green) line is for $n=9$, the
dashed (blue) line is for $n=13$, and the solid (red) line is for
$n=20$. (b) FCC lattices. The dotted (brown) line is are for $n=4$,
the dotdashed (green) line is for $n=6$, the dashed (blue) line is
for $n=8$, and the solid (red) line is for $n=13$. $n$ is the number
of lattice points on the edge of the cubic lattice, from which we
cut the inscribed sphere.} \label{fig:contri}
\end{figure}

Four different finite sizes for SC and FCC lattices are shown in
Fig.~\ref{fig:contri} respectively. Noted that there is a cross
point in both parts of the figure, i.e.~$\omega\approx2.6$ for SC
and $\omega\approx4.5$ for FCC, so the eigenfrequencies are divided
into two parts naturally: low and high eigenfrequencies. For the
smallest size (the dotted lines) shown in Fig.~\ref{fig:contri},
most contributions are from the low eigenfrequencies and the line
fluctuates strongly in the low eigenfrequency range. With the
increase of the size, the contributions from the high
eigenfrequencies increase, but the increase becomes smaller and
smaller; the contributions from the low eigenfrequencies get smaller
with increasing size and also there is a smaller and smaller
decrease, but the change is much bigger than that of the high
eigenfrequencies. Apparently, a peak corresponding to the peak at
$\omega\approx2.4$ for SC and at $\omega\approx3.5$ for FCC in
Fig.~\ref{fig:dos} respectively appears in the lines (the dotdashed
line and the dashed line) for the two middle sizes. Similarly, a
peak corresponding to the peak at $\omega\approx4.6$ for SC and at
$\omega\approx7.5$ for FCC appears in the line (the solid line) for
the largest size in Fig.~\ref{fig:contri}, which will be apparent if
we plot it alone. So the line of contribution becomes more similar
to the line of density of phonon modes, when the size becomes
larger. The contributions almost vanish at the highest
eigenfrequencies. So the low eigenfrequencies dominate the melting
for relatively small nanoparticles and they can be excited at low
temperature. But for relatively large nanoparticles, the high
eigenfrequencies nearly give the same contributions to the melting.
That is why the nanoparticle melting temperature decreases with
decreasing size.

We can show that all the phonon modes with small mean neighbor
relative deviations, $\langle (\bm{u}^i-\bm{u}^j)^2\rangle$, belong
to the low eigenfrequency part. We call these acoustic-like phonon
modes. Most phonon modes with large mean neighbor relative
deviations are within the high eigenfrequency part, and we call them
optical-like phonon modes. In the acoustic-like modes, neighbor
atoms tend to vibrate in the same direction similar to the acoustic
modes in the diatomic linear chain; while in the optical-like modes,
they prefer to vibrate in the opposite direction. So with the same
thermal vibration energy, the acoustic-like modes can result in
larger atomic mean-squared thermal vibration amplitude, but the
optical-like modes can not do that because they have large potential
energy when the neighbor atoms move to each other with small
displacements.

\begin{figure}
\includegraphics[width=8cm]{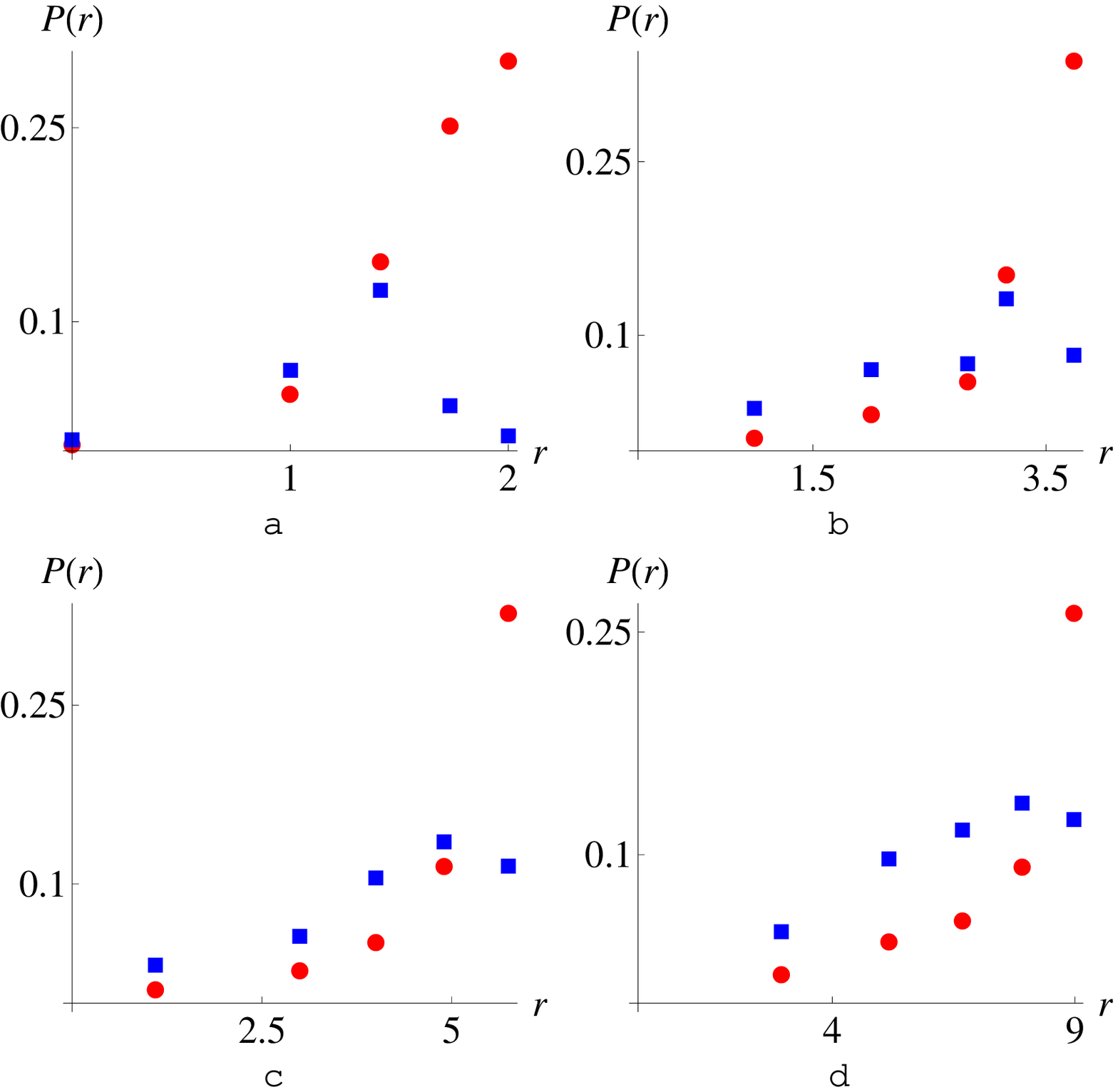}
\caption{(Color online)Contributions from the low and high
eigenfrequencies from five layers with mean radius $r$ from the
center of the inscribed sphere for SC lattices. The circles (red)
are for low eigenfrequencies and the squares (blue) are for high
eigenfrequencies. (a) $n=5$ (b) $n=9$ (c) $n=13$ (d) $n=20$. $n$ is
the number of lattice points on the edge of the cubic lattice, from
which we cut the inscribed sphere.} \label{fig:scr}
\includegraphics[width=8cm]{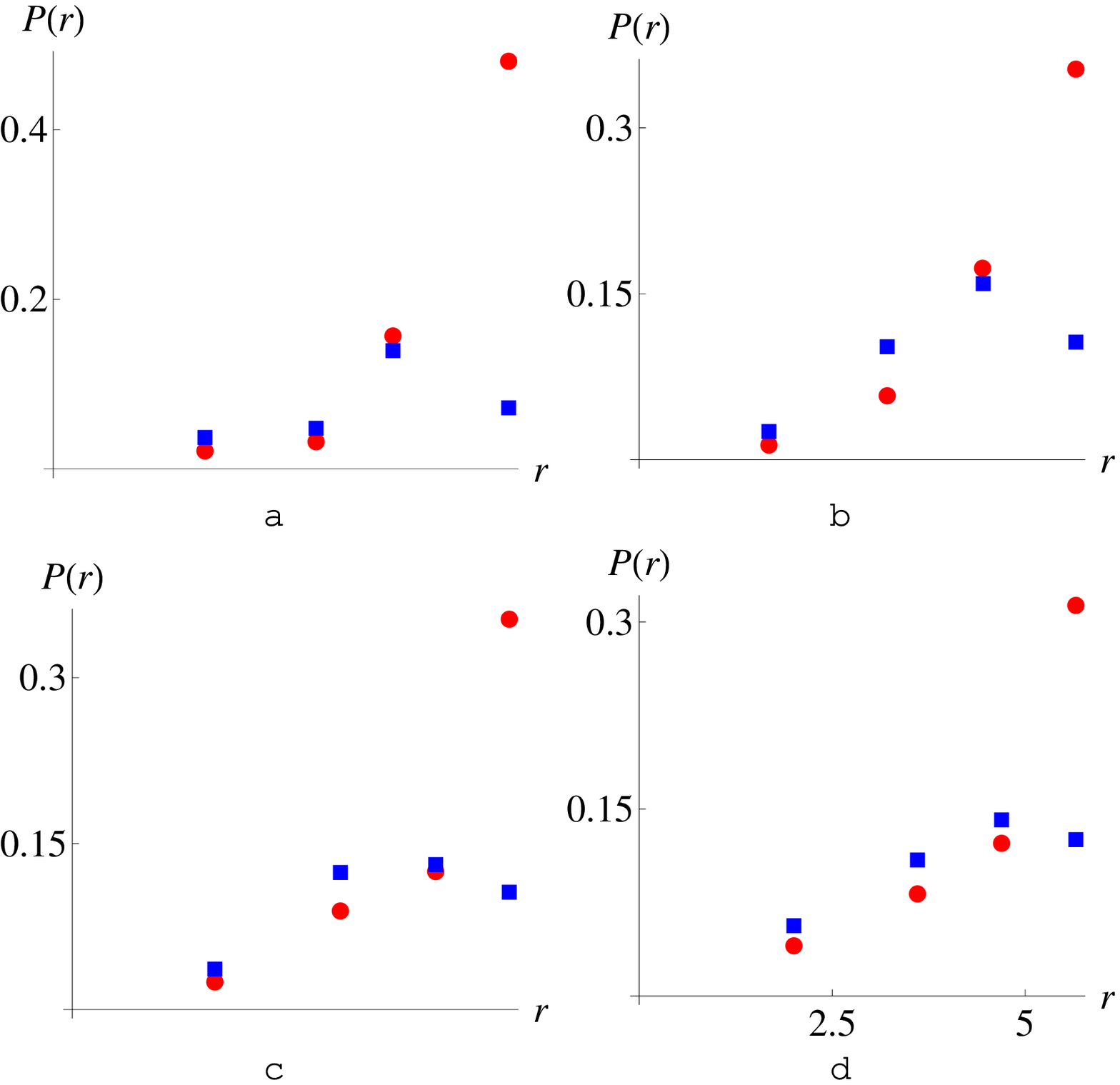}
\caption{(Color online)Contributions from the low and high
eigenfrequencies from four layers with mean radius $r$ from the
center of the inscribed sphere for FCC lattices. The circles (red)
are for the low eigenfrequencies and the squares (blue) are for the
high eigenfrequencies. (a) $n=4$ (b) $n=6$ (c) $n=9$ (d) $n=13$. $n$
is the number of lattice points on the edge of the cubic lattice,
from which we cut the inscribed sphere.} \label{fig:fccr}
\end{figure}

Then we divide the nanoparticle into five layers for SC and four
layers for FCC, each layer having a mean radius $r$ from the center
of the inscribed sphere, and plot the contributions from the low and
high eigenfrequencies respectively versus $r$
(i.e.~$P(r)=\frac{1}{\langle u^2\rangle_{tot}}\sum''\langle
u_{i\alpha} ^2 \rangle_{\Delta\omega}$, where $\sum''$ means the sum
is taken over those $i,\alpha$ whose corresponding atoms are in the
same layer and $\langle u_{i\alpha} ^2 \rangle_{\Delta\omega}$ means
the sum of $k$ in the expression of $\langle u_{i\alpha} ^2 \rangle$
is taken over those $k$ whose corresponding $\omega_k$ belong to the
low or high eigenfrequencies; $\langle u^2\rangle_{tot}$ appears for
normalizaion) in Fig.~\ref{fig:scr} for SC and Fig.~\ref{fig:fccr}
for FCC, both with four different finite sizes respectively. Here
all $\langle u_{i\alpha} ^2 \rangle$ are considered no matter
whether they contribute to the melting or not. This is different
from Fig.~\ref{fig:contri}.

Apparently, the low eigenfrequencies vibrate the atoms near the
surface most, and they have nearly no effect on the inner atoms. It
means surface-premelting, because the low eigenfrequencies are
excited at low temperature and they mainly vibrate the atoms near
the surface. This behavior should be a general property in the model
similar to ours. The high eigenfrequencies vibrate the atoms in the
middle when the size is relatively small, and they vibrate the atoms
all but the inner ones when the size becomes large. So when most
atoms are near the surface, the low eigenfrequencies is sufficient
for melting because of their low excitation temperature, and when
the size becomes large, more and more atoms are in the middle, the
high eigenfrequencies have to be excited at high temperature for
melting.

\begin{figure}
\includegraphics[width=8cm]{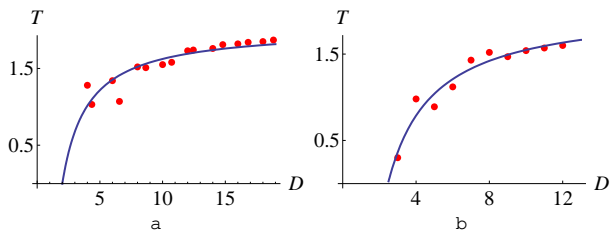}
\caption{(Color online)The relation between nanoparticle melting
temperature and nanoparticle size. (a) SC lattice. $n$ is from 5 to
20. The fit line is $T=2.03- 4.06536/D$. (b) FCC lattice. $n$ is
from 4 to 13. The fit line is $T=2.05495-5.07076/D$.}
\label{fig:fit}
\end{figure}

The relation between nanoparticle melting temperature and
nanoparticle size are shown in Fig.~\ref{fig:fit}. The data
calculated for SC and FCC lattices are fitted by $T_{mb}-A'/D$. We
do not use $1-A/D$ because we have no idea on the bulk melting
temperature $T_{mb}$ for our model. It can be seen that the linear
decrease of the nanoparticle melting temperature with decreasing
size is reproduced for the two lattices. And the nanoparticle
melting temperature fluctuates stronger with decreasing size. We
think the reason is that when the size decreases, the density of
phonon modes gets irregular, so it is uneasy to find a regular line
to fit these small sizes. $A=A'/T_{mb}$ is approximately 2.0 for SC
and 2.5 for FCC. The difference comes from the fluctuation of the
nanoparticle melting temperature of small sizes. If we only consider
the large sizes calculated in our model, SC and FCC will have nearly
the same $A$ in our calculation ($A=2.2$ for SC and $A=2.3$ for
FCC).

\section{CONCLUSIONS}
In this paper, we numerically calculate the phonon spectra for two
types of three-dimensional lattices (SC and FCC), in the framework
of a very simple microscopic model with a spring interatomic
interaction within the harmonic approximation. Each spatial
component of the mean-squared thermal vibration amplitude of each
atom can be obtained during the calculation, and we are able to
introduce a statistical melting criterion similar to the Lindemann
criterion. Within this criterion, the linear decrease of the
nanoparticle melting temperature with decreasing nanoparticle size
and irregular variations on a fine scale are reproduced by using our
simple model. And we found that the eigenfrequencies are naturally
divided into two parts, the low eigenfrequency part and high
eigenfrequency part, each of which ought to include all the
acoustic-like and most of the optical-like modes respectively. The
low eigenfrequency part played a major role in the melting of small
nanoparticles, which resulted in the depression of the melting
temperature. The atoms mostly near the surface are mainly vibrated
in these modes, that is the so-called surface-premelting. This
should be a general feature when considering the mean-squared
thermal vibration amplitude of each atom independently. In this
paper, we only consider the case of monoatomic nanoparticles, and
the case of diatomic nanoparticles is under investigation. Moreover,
we are planning to employ a more realistic interatomic interaction
(including the electron-phonon coupling) and introduce relaxation of
the structure in the further study.

\end{document}